\documentclass[conference]{IEEEtran}
\IEEEoverridecommandlockouts

\usepackage{cite}
\usepackage[T1]{fontenc}
\usepackage[utf8]{inputenc}
\usepackage{amsmath,amssymb}
\usepackage{url}
\usepackage{algorithmic}
\usepackage{graphicx}
\usepackage{tabularx}
\newcolumntype{C}{>{\centering\arraybackslash}X}

\usepackage{textcomp}
\usepackage{xcolor}
\usepackage{subcaption}
\def\BibTeX{{\rm B\kern-.05em{\sc i\kern-.025em b}\kern-.08em
    T\kern-.1667em\lower.7ex\hbox{E}\kern-.125emX}}

\begin{document}

\title{Risk-based Design for Sustainability in Cloud Systems: Insights from an Experts' Survey \\
}

\author{\IEEEauthorblockN{Maria Voreakou, Cleopatra Bardaki, George Kousiouris, Mara Nikolaidou}
\IEEEauthorblockA{\textit{Department of Informatics and Telematics} \\
\textit{Harokopio University of Athens}\\
Athens, Greece \\
voreakou@hua.gr, cleobar@hua.gr, gkousiou@hua.gr, mara@hua.gr}


}

\maketitle

\begin{abstract}
Cloud Systems' Sustainability is critical in Cloud Computing, especially with the growing demand in many industries. Sustainability risks in Cloud Computing can be tricky, mostly because of system complexity and their impact on performance. Thus, this research focuses on Risk-based design (RBD) and how it can support the early identification of possible risks for Cloud System Sustainability, as well as respective mitigation strategies of each risk. In order to successfully identify risks of Cloud System Sustainability, an Expert Survey is conducted including experts with different roles from different industries, to identify possible sustainability risks of a Cloud System on all different levels. Thematic analysis of the responses resulted in a categorization of risks, as well as in the identification of mitigation strategies and factors affecting each risk. Such findings can be helpful for researchers and practitioners that utilize RBD when building sustainable Cloud systems.

\end{abstract}


\section{Introduction}
Enterprises have embraced cloud systems primarily because of their advantages, such as expandability, transmutability, and other economic based factors. As Cloud Systems need to provide high availability and reliability, energy costs on Cloud Systems arise \cite{gill2018taxonomy}. However, this increasing dependence on cloud infrastructure raises sustainability challenges, especially regarding sustainable economic performance, ecological sustainability, and internal efficiency. Given the fact that processing and storing data further impacts the sustainability of Cloud Systems, and is directly connected with the large energy consumption, there is a lot of research around Cloud System Sustainability \cite{mastelic2014cloud} \cite{ficco2016economic}. Risk-Based Design (RBD) is one of the vital methods which helps overcome these issues, as it allows risk to be taken into consideration, as appropriate, over the lifetime of a Cloud System. 

RBD is defined as a reliability-based design methodology and uses \textit{risk} as an indicator to set the functions' criteria such as the upper limit of the probability of failure in structural design. The use of \textit{risk} as an outcome indicator, increases the universality and transparency of the evaluation criteria \cite{makoto2022fundamentals}. 

Risk-based design has been applied across various sectors including energy systems, engineering education, construction, cloud computing, and more. Each approach demonstrates how RBD can be adapted to unique industries to mitigate every risk and improve their performance, security, but also sustainability of the systems.

In our case, risk in Cloud Systems is defined as the potential for an unwanted or adverse event to occur, along with the likelihood and impact of that event \cite{paper5} \cite{paper6}. These risks can arise from a variety of factors, including technical failures, human errors, environmental challenges, and economic or legal shifts \cite{paper1} \cite{paper4}. Managing risks at the initial stages of potential cloud technological applications up to an acceptable level is what RBD is all about. It involves controlling and mitigating threats in cloud systems.

The key objective of our research is to identify risks and mitigation practices while applying RBD in order to build a Sustainable Cloud System. We perceive Cloud System Sustainability as the conformance of its operation with environmental, economical and governance principles enabling their operation in the future and eliminating related threats. To this end, through RBD, a) we identify any possible related risks, in these three categories, b) we recognize methods for System Resilience and Optimization, and c) we identify if proactive Risk Management can be used to avoid Related Risks and/or provide measures/ mitigation strategies to deal with the Risks if needed.

We conduct an Expert Survey, interviewing a group of Experts involved in Cloud Computing environments, and continue with analyzing the data collected to understand the risks the experts recognize. The ultimate goal of this paper is to identify categories of risks that may appear in the whole lifecycle of a Cloud System project, as well as means to overcome them in order to ensure sustainability. Our findings are empirical, based on experts; and provide valuable insights on risks and risk management to researchers and designers that want to develop and operate sustainable cloud systems.

The rest of this paper is structured as follows: In Section II we investigate the literature on RBD and sustainability of cloud systems. In Section III we present the results of RBD assessment and expert evaluations that we have conducted, while also discussing competitive threats to cloud systems' sustainability. In Section IV we present the results and in Section V conclude this research and present possible future work.

\section{Related Work}

A critical analysis report on sustainable energy system planning \cite{paper1} has employed techniques like Mean-Variance Portfolio Analysis along with the use of Monte Carlo Simulation. The aim of these approaches was to design optimal efficient energy systems which would withstand uncertainties like changing demand and fuel prices. With the help of RBD, these systems were also made economical and able to withstand a variety of risks in the future. 

In the context of process safety and sustainability, in their paper \cite{paper2} the authors discussed how RBD is used to teach future engineers on how to manage complex risks in engineering systems. The study concluded that by introducing sustainability principles into risk based education (RBD), RBD prepares engineers for the future and the design of systems with less risk to the society.


The authors of \cite{paper4} made a systematic risk assessment by using the Risk-based design on the life-cycle of cloud systems. They were able to identify risks earlier, such as security, resource management, and scalability risks, and as a result, this approach ensured that the cloud services were built in a sustainable and resilient way at their core.

Authors of \cite{abdul2017risk} presented a risk management approach in order to achieve a Sustainable Cloud Migration. To define this approach, they consider 4 dimensions of sustainability, in order to better determine the Cloud viability for the business context.

A risk perception study, described in \cite{paper5}, was executed with a group of Swiss companies, which focused on managing cloud computing risks such as data security, regulatory compliance, and disaster recovery. The study demonstrated how a risk-based approach could tailor cloud services to meet specific needs, ensuring that the systems were secure and sustainable.

In \cite{paper6} the authors talk about a risk management framework for cloud-based services innovation. By addressing the process risks, services, and technology, the authors ensured that cloud innovations were sustainable and aligned with business goals, mitigating both technological and operational risks. 

In \cite{paper7} the authors conduct a review focused on the integration of security risk management into cloud-based business processes, for which they applied RBD to ensure a secure and efficient process execution. The study emphasized proactive risk management across the life-cycle of cloud systems to maintain both sustainability and security.

The authors of \cite{paper8} introduced a structured risk-based framework which included Risk Assessment as a Service (RAaaS) to manage risks in their cloud infrastructures. The framework integrated risk management at each phase of the system design, from inception to monitoring, ensuring that the systems were sustainable and secure.

The work of \cite{paper9} addresses sustainable construction, and used RBD to assess the environmental impacts of construction projects. By balancing time, cost, and sustainability, this study developed a framework that prioritized environmental sustainability while managing risks such as resource scarcity and pollution.

The authors of \cite{paper10} combined the Delphi method and Analytic Network Process to rank cloud risks and framework mitigation strategies. They introduced a formal decision-making model integrating costs and risks, using Reliability Block Diagrams (RBD) to quantify confidentiality, integrity, and availability. This structured approach guides cloud provider selection while ensuring financial, security, and operational sustainability.

A risk-based offloading model for mobile cloud computing is proposed in \cite{paper12}. By evaluating both offloading benefits and risks (e.g. privacy breaches and network reliability), the study used RBD to optimize offloading decisions for energy savings and improved performance. Similarly, authors of \cite{paper13} built an access control architecture with RBD principles, and they used a dynamic risk engine to process access requests in real time. This flexible, risk-based approach allowed the cloud systems to dynamically respond to security risks, by enhancing their sustainability and scalability.


While most of the aforementioned works apply RBD in Cloud Systems, each of them focuses on some aspects of Cloud Computing. For example, studies like \cite{abdul2017risk} \cite{paper6} \cite{paper4} were able to identify exclusively technical risks (resource management, etc), while others like \cite{paper7} were focused only in Business Processes. In our study, we want to identify possible risks in the complete project lifecycle of designing a sustainable cloud System, including the inception phase and all project stages until completion, with the ultimate target to be that the cloud System is sustainable. In short, we adopt a holistic view of a Cloud system design project and look for risks pertaining to each project phase.

\section{RBD Assessment of Cloud System Sustainability}


The experts in the survey consisted of 9 IT experts from different fields who are in leading positions or manage main products of each company. 55\% are Senior Software Engineers, 12\% Senior Data Analysts, and 33\% System's Engineers \& IT Architects. The average time of completing the survey was 20 minutes. We used Google Forms as the survey tool for our study. There were 8 main Questions which included open questions with a paragraph answer, Multiple-choice Questions, Short Questions, as well as Multiple-choice Grid Questions. In addition, the survey had an introduction to RBD, to make sure all the participants know the definitions of RBD and System Sustainability, as well as the goal of the survey. The survey itself can be found in the GitHub Repository of this research \cite{githubMariaVoreakou}, as well as the dataset including the questions and the answers of all the participants. 

We explored sustainability, environmental, economical and governance risks of Cloud Systems in different levels, where risks are categorized in different levels \cite{abdul2017risk}. Similarly, in our study we split the Risk Levels into \textit{Application}, \textit{System}, \textit{System of Systems}, \& \textit{Business} Levels in order to define different responsibilities, and identify related sustainability risks in each Level. The risk levels are explained below.

\begin{enumerate}
\item 
    \textbf{Application Level}: At this level, we have risks that include resource reduction like thread or memory to mention a few, or other alterations such as transferring a service to a different programming language. Such risks would lead to problems like the application exhibiting reduced performance, bugs or even inefficient coding for engineers unfamiliar with the new tool-set. 
    \item 
    
    \textbf{System Level}: The risks in this level affect the entire system and include resource cuts in computational resources on a higher level, like for example less available pods in a cloud management environment like \textit{Kubernetes}, reduced hypervisor performance and so on. Risks like these are likely to cause a slower system and even result in a system being shut down resulting in delays for the service being offered. 
    \item 
    
    \textbf{System of Systems (SoS) Level}: At this level the risks arise from the interactions between various systems. Changes at the application or system level can create instability in interconnected systems, reducing the overall system availability. This level can have similar risks to Application and System Levels.
    \item 
    
    \textbf{Business Level:} Business-related risks include the resistance to adopt new technologies or sustainability practices, and are directly connected with all the people in a company to different levels. These kind of risks may cause investments in sustainability that may not yield immediate financial returns, and for example there could be a shortage of skilled engineers to implement necessary changes.
\end{enumerate}

\begin{figure}
    \centering
    \includegraphics[width=1\linewidth]{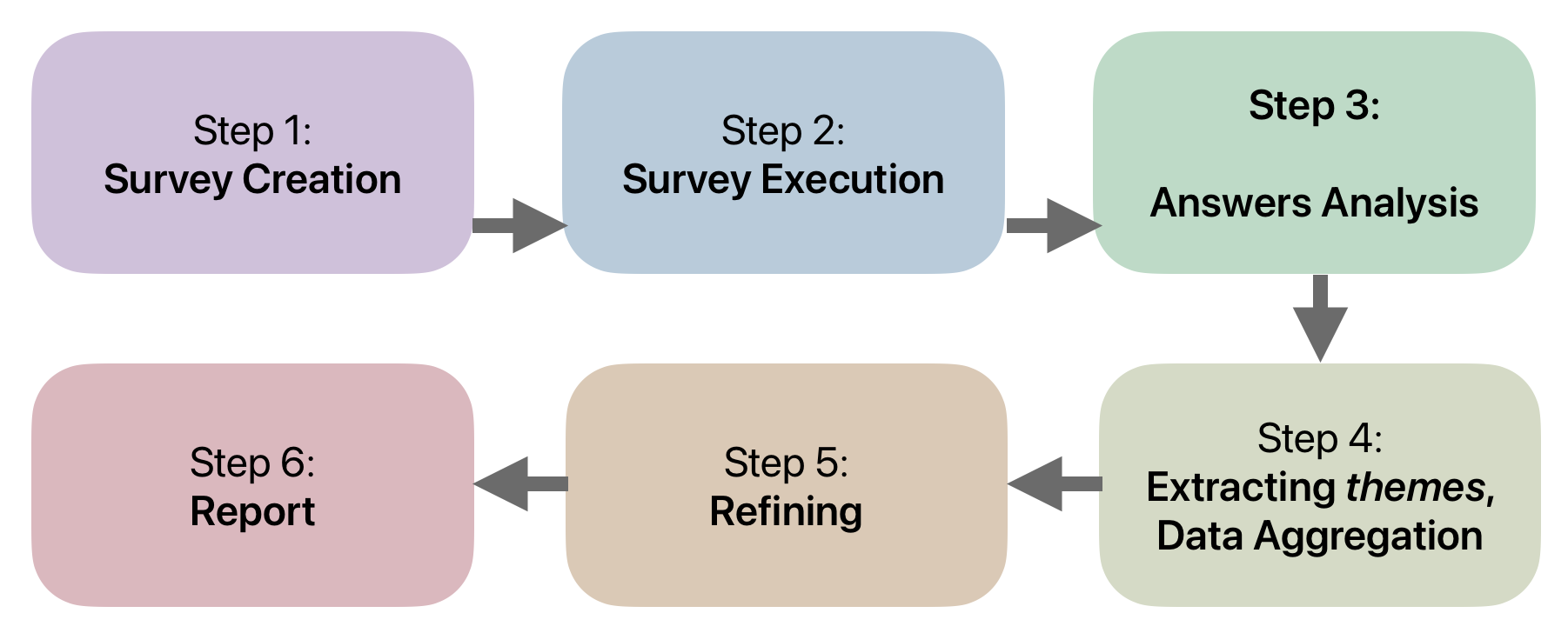}
    \caption{Analysis of Expert Survey's results}
    \label{fig:thematic-analysis}
\end{figure}

Given that this expert survey included open questions, in order to have a more complete view of each participant's opinion, we have executed multiple steps on analysis to come up with a more generalized categorization of risks across all \textit{Risk Levels}, and come up with the final outcome.
We have used \textit{Thematic Analysis} \cite{thematicanalysis} (step 4 in Fig. \ref{fig:thematic-analysis}) to categorized risks identified. Thematic analysis fits our research, since it categorizes qualitative data into \textit{themes} based on common patterns and provides systematic procedures to generate \textit{codes} from qualitative data. Codes are small units of analysis that capture interesting features of the data relevant to the research questions provided to the Survey. By aggregating Codes as building blocks, \textit{themes} are created, which are either patterns or a more generic concept or core idea. In our work, \textit{themes} are represented as \textit{Cycles}, while \textit{Codes} are the categorized \textit{Risks}.

\begin{figure}
    \centering
    \includegraphics[width=1\linewidth]{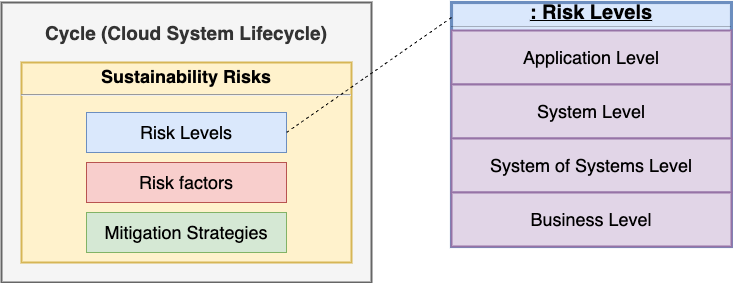}
    \caption{Cycle Entities Legend}
    \label{fig:sysml-cycles}
\end{figure}

On the other hand, we categorized the \textit{Risks} into 4 \textit{Cycles} as proposed in \cite{abdul2017risk}. The \textit{Cycles} are responsible for each stage of the Cloud System development, while in each \textit{Cycle} multiple \textit{Sustainability Risks} can be incorporated. 
Finally, during the survey, the experts responded to each risk with \textit{Mitigation Strategies} and \textit{Risk Factors} related to sustainability issues. The expert survey was finally structured as follows: \textit{Cycles}, \textit{Risks}, \textit{Risk Levels}, \textit{Mitigation Strategies}, and \textit{Risk Factors}, and the fully relation of each entity is explained in Fig. \ref{fig:sysml-cycles}.

\section{Results}

\begin{table*}[t]
    \begin{minipage}{0.49\textwidth}
        \centering
        \refstepcounter{table}
        \label{table_cycle}
        
        {\small\scshape Table \thetable}\\
        {\small Risks \& Mitigation Strategies identified by Cycle}\\[1ex]
        
        \begin{tabularx}{\linewidth}{|X|C|}
            \hline
            \multicolumn{1}{|c|}{\textbf{Cycle}} & \textbf{Percentage} \\ \hline
            C1: Identification & 21.4\% \\ \hline
            C2: Analysis       & 21.4\% \\ \hline
            C3: Prioritization & 28.6\% \\ \hline
            C4: Decisions      & 28.6\% \\ \hline
        \end{tabularx}
    \end{minipage}
    \hfill 
    \begin{minipage}{0.49\textwidth}
        \centering
        \refstepcounter{table}
        \label{table_level}
        
        {\small\scshape Table \thetable}\\
        {\small Risks \& Mitigation Strategies identified by Level}\\[1ex]
        
        \begin{tabularx}{\linewidth}{|X|C|}
            \hline
            \multicolumn{1}{|c|}{\textbf{Risk Level}} & \textbf{Percentage} \\ \hline
            Application Level & 37.5\% \\ \hline
            System Level      & 18.8\% \\ \hline
            System of Systems & 25\%   \\ \hline
            Business Level    & 18.8\% \\ \hline
        \end{tabularx}
    \end{minipage}
\end{table*}

The usage of \textit{thematic analysis} methodology gave us the following data aggregation as shown in Tables \ref{table_cycle} and \ref{table_level}. 
In Table \ref{table_level}, we present the results of the participant responses in percentage of the risks identified per Risk Level, while in Table \ref{table_cycle} we show the percentage of all the risks identified by the participants categorized into 4 \textit{Cycles}, which all the cycles together define the stages of a project lifecycle. 

\begin{figure*}[t]
\centerline{
\includegraphics[width=1\linewidth]{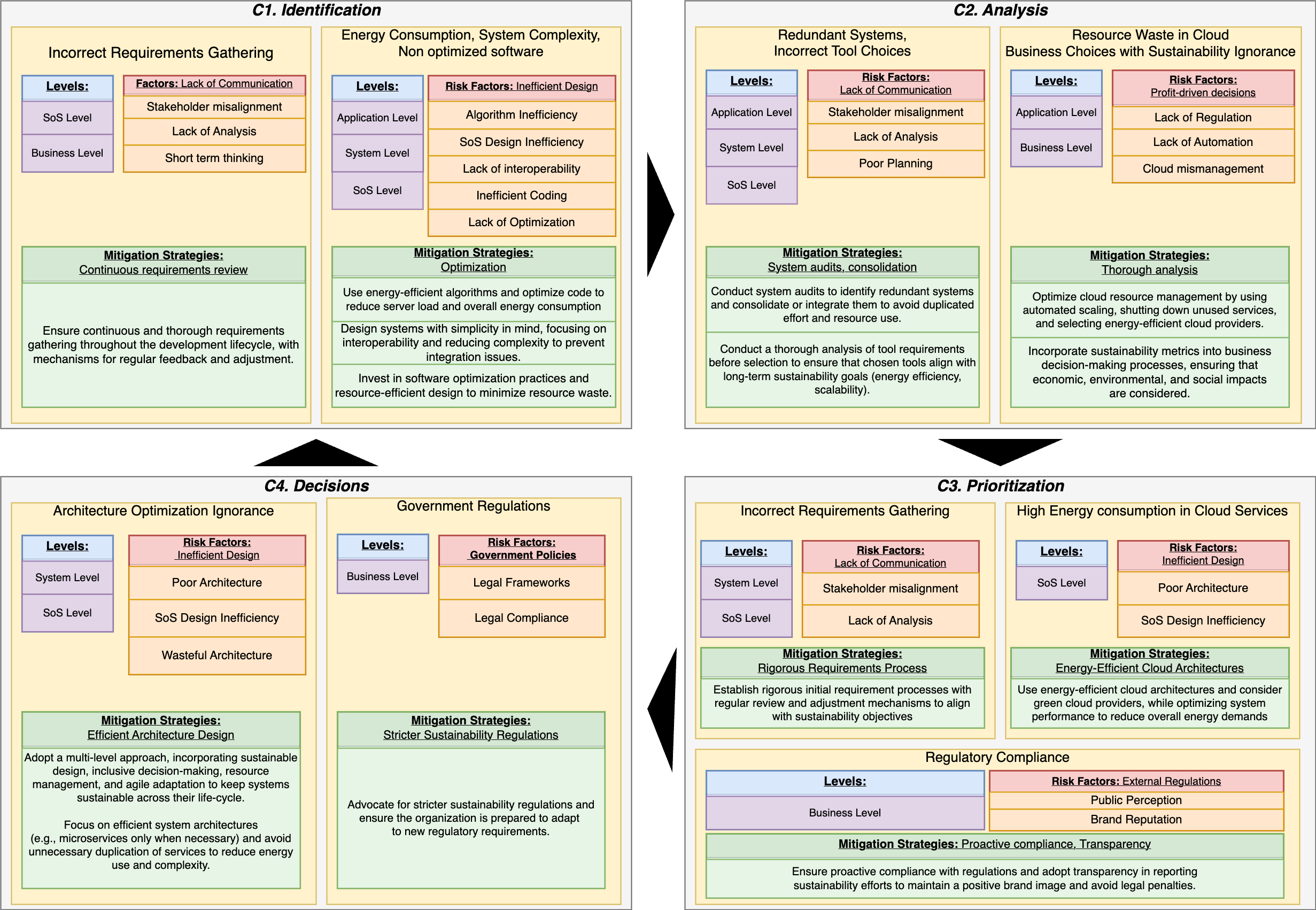}
}
\caption{RBD in Cloud System Sustainability: Expert Survey Overview in Cycles.}
\label{fig:overview-cycles}
\end{figure*}

In Fig. \ref{fig:overview-cycles} we present the outcome of this research, categorized it in 4 Cycles we have concluded, including the Risks, Risk factors, and Mitigation Strategies per Cycle. In Each Cycle, multiple \textit{Risks} can be included, and in the next section each Risk will be described further together with their mitigation strategies and Risk factors or influences. 

\subsection{Cycles Definition} 

\textbf{\textit{C1: Identification}}: This is the first \textit{Cycle}, where we identify risks at a very primal stage. We do not yet have clarity of all the Risks involved in a project lifecycle, and we can have Risks from all the different Risk Levels. 

\textbf{\textit{C2: Analysis}}: Usually, during this phase, we find Risks related to lower Risk Levels which means that risks are closer to the applications and the systems. 

\textbf{\textit{C3: Prioritization}}: In this \textit{Cycle}, we find more \textit{Risks} from Business and System of Systems \textit{Risk Levels}. During this phase the project development is in a more stable stage, the system or the application is live for some time, and during some evaluations, people can explore future risks. 

\textbf{\textit{C4: Decisions}}: This is the last phase of a project lifecycle, and 2 main \textit{Risks} are identified in this \textit{Cycle}. As it is the final phase of a project's lifecycle, in this stage decisions are made based on the previous \textit{Cycle} Risks.

\subsection{Risks identified}
A Risk can be present in one or multiple Cycles, depending on the complexity of the Risk itself. Below, an additional elaboration per Cycle is done, and we also introduce the factors and the Risk Levels each Risk is applied to.

\textbf{\textit{C1 Identification}}: 
The first Risk factor that we conclude is the \textit{\textbf{Lack of Communication}}. Lack of Communication influences the \textit{Incorrect Requirements Gathering} Risk, which also includes the \textit{Stakeholder misalignment}, \textit{Lack of Analysis}, and \textit{Short Term Thinking}. This factor is intangible as it is based on people's thinking and sometimes the education of the whole team involved. Lack of Communication is also influenced by the multi-international environment that now most of the companies have. 

Another Risk factor is the \textit{\textbf{Inefficient Design}} which influences more than one Risks (\textit{Energy Consumption}, \textit{System Complexity}, \textit{Non Optimized software}). More specifically, the aforementioned Risks are influenced by the \textit{Algorithm inefficiency} and the \textit{SoS design inefficiency}, \textit{poor design}, \textit{lack of interoperability}, as well as \textit{Inefficient Coding} and \textit{Lack of Optimization}. Some of the aforementioned topics are tangible and some intangible. For topics such as \textit{Inefficient Coding} or \textit{Algorithm inefficiency}, tools are created to measure how efficient the code is. Although these tools might not be 100\% accurate, they can help teams be more productive and focus on more important tasks. All the other topics are intangible, as for example the \textit{Poor design} cannot be measured with precision, but an Engineer can employ the use of existing design patterns to improve a poor design. The expert Group provided that a mitigation strategy for all of the above topics can be the \textit{Usage of Energy efficient algorithms}, \textit{Simplified System Design}, and \textit{Software Optimization}.

\textbf{\textit{C2 Analysis}}: 
\textit{\textbf{Lack of Communication}} is again a factor for the Analysis Cycle.
The expert group answered that two main Risks are influenced by this factor: \textit{Redundant Systems} and \textit{Incorrect tool choices}. These Risks can be found at 3 different Risk Levels, \textit{Application}, \textit{System} and \textit{System of Systems} Level. Lack of Communication also includes \textit{Stakeholder Misalignment}, \textit{Lack of Analysis} and \textit{Poor Planning}. Again we see that some influences might be tangible and some intangible. The Expert group answered that these risks can be mitigated by having systematic \textit{System Audits}, and \textit{System Consolidation} in order to avoid redundant systems. They also mentioned that \textit{Thorough Tool Analysis} can mitigate the incorrect tool choices and help in long-term thinking.

Another factor that influences risks in this cycle is \textit{\textbf{Profit Driven Decisions}}. More specifically, it influences two main Risks which are \textit{Resource Waste in Cloud}, and \textit{Business choices with Sustainability Ignorance}. These Risks can be found in \textit{Application} and \textit{Business} Levels. Additionally, this factor includes influences such as \textit{Lack of Regulation}, \textit{Lack of Automation} and \textit{Cloud Mismanagement}. In order to mitigate them, the expert group answered that \textit{Automated Scaling}, and \textit{Shutdown of unused Resources} can help improve the Cloud Mismanagement and automation. They also mentioned that \textit{Sustainability Metrics} can be incorporated into business decision-making processes, and as a result they will ensure that economic, environmental \& social impacts will be considered. All the influences are tangible and can be managed through automations and implementation of metrics in the right areas of each environment.

\textbf{\textit{C3: Prioritization}}:
\textit{\textbf{Lack of Communication}} and \textit{Incorrect Requirements Gathering} reappear as risks in this cycle. Incorrect Requirements Gathering is additionally  associated with the \textit{Business} and \textit{System of Systems} Risk Levels. The expert group answered that in order to mitigate the aforementioned risk, a \textit{Rigorous Requirement Process} is necessary, with regular review and adjustment mechanisms to align with sustainability objectives. This factor can be tangible if a process methodology with district steps in order to measure the factor and how much it was improved is created or provided.

\textit{\textbf{Inefficient Design}} is another factor that we found in this Cycle.
This factor influences the \textit{High Energy Consumption in Cloud services} Risk, and can be associated with the \textit{System of Systems} Level. Inefficient design also includes \textit{Poor Architecture}, and \textit{SoS Design Inefficiency}. The expert group answered that in order to mitigate this risk, we need \textit{Energy-Efficient Cloud Architectures} and to consider Green Cloud providers, while optimization of System performance will reduce the overall energy demands. In order to measure this factor and be tangible, it is required to have transparency of the Cloud Provider and how the Cloud Architecture can be Green.

Another major Risk factor in this Cycle is the \textit{\textbf{External Regulations}}.
External Regulations influence the \textit{Regulatory Compliance} Risk, and this Risk is found only in the \textit{Business} Level. External Regulations include as additional influences the \textit{Public Perception} and \textit{Brand reputation}. The expert group answered that in order to mitigate the Regulatory Compliance, we have to ensure a proactive process for compliance with regulations, and adopt transparency to report sustainability efforts. They also mentioned that this mitigation strategy will help to maintain a positive brand image and avoid legal penalties. In order to measure this factor and be tangible, it is required to have a discrete process methodology applied, as well as to be required to all the Business processes and decisions.

\textbf{\textit{C4:Decisions}}:
\textit{\textbf{Inefficient Design}} is included in the last project lifecycle, and influences one main Risk, the \textit{Architecture Optimization ignorance} which can be found in \textit{System} and \textit{SoS} Risk Levels. Additionally to this factor are included the \textit{Poor Architecture}, \textit{SoS Design Inefficiency}, and \textit{Wasteful Architecture}. The expert group answered that in order to mitigate Architecture Optimization ignorance, we need to focus on efficient system architectures with an example of using microservices only if necessary. The expert group also mentions that we have to avoid unnecessary duplication of services to reduce energy and system complexity. This risk can be tangible only if a methodology to compare system performances is used, and by setting up an overall monitoring system with metrics collected from all the microservices of the cloud system inspected, while also looking at the systems' behaviors. 

Another factor included in this Cycle is \textit{\textbf{Government Policies}} which actively influences the main Risk \textit{Government Regulations}, and can be found only on the \textit{Business} Risk Level. Additionally to this factor, influences such as \textit{Legal Frameworks} and \textit{Legal Compliance} are also major for this risk. The expert group answered that in order to mitigate those Policies, we have to advocate for stricter sustainability regulations, as well as to ensure the organization is prepared to adapt to new regulatory requirements. This risk category is intangible as it is not clear how to measure it, however if a process methodology is going to be applied, then it can be easier for companies to understand their progress to this mitigation strategy.

\section{Overall Findings}

Overall, the expert group emphasized that Cloud Systems need to be sustainable targeting  Environmental, Economic and Governance priorities. They mentioned practices to assess cloud systems in order to reduce their environmental footprint, manage waste of energy consumption, redundancy of microservices, as well as how to improve system design in order to avoid energy waste.

\subsubsection{Environmental Sustainability}

85.7\% of respondents emphasized that Business level risks, such as prioritizing profits over sustainability, are particularly concerning. At the System-of-Systems (SoS) level, 71.4\% highlighted risks stemming from inefficient inter-system dependencies, such as redundant systems leading to resource waste. At the System level, 57.1\% of the expert group pointed to risks like inefficient resource management and energy usage. A significant 42.9\% considered the Application level risks (e.g. energy consumption and maintenance costs) to be of importance.

At the Application level, 42.9\% of the respondents mentioned the risk of high-traffic applications with inefficient algorithms leading to excessive energy use. At the system level, 57.1\% referred to outdated data centers as major environmental risks. For system-of-systems risks, 71.4\% mentioned the inefficiency of interconnected energy systems in smart cities.

\subsubsection{Economic Sustainability}

Targeting economic sustainability, the expert group focused on enhancing System Resilience and Optimization,  prioritizing System Integrity together with the System Scalability as the most important issues. The expert group's answers were focused on general Systems Optimizations including the reduction of redundant systems or non optimized systems by finding ways to regulate methodologies and processes. They also mentioned that there is a need for Sustainability Encouragement and Awareness in order to apply these Architectural Optimizations and Business Practices in practice. 28.6\% focused on profit-driven business practices, arguing that short-term profitability often undermines long-term sustainability.

\subsubsection{Governance Sustainability}

42.9\% of the respondents identified SoS risks as the most critical due to the complexity of managing interdependent systems. 28.6\% considered the collection of incorrect requirements as critical, highlighting that errors made in the early stages of a project can cascade throughout its lifecycle. Design choices were seen as a key factor by 42.9\% of the respondents, with operational practices also influencing risk severity for 28.6\%. External regulations were deemed to have the strongest influence by 57.1\% of the expert group, highlighting the importance of regulatory frameworks in driving sustainability.

The respondents cautioned against misusing sustainability indicators and shared their views on the importance of transparency in sustainability initiatives. Some claimed that, in addition to managerial strategies, more comprehensive political solutions are needed to properly handle sustainability concerns.

Finally, to guarantee a Proactive Risk Management, the expert group's answers were focused on the Identification and the Mitigation of the risks mostly in the early stages of the Cloud System lifecycle. They also mentioned that it is important to develop methodologies or frameworks for early risk assessment, in order to make the risks tangible.

\section{Conclusion}
RBD can provide a complete approach for resolving the risks during a sustainable cloud system project. The experts survey provided great insights into the challenges and risk management approaches, underlining the need to comprehensively address all risk levels. Given that, organizations must adopt strategies that ensure cloud system sustainability. This requires integrating technical and operational risk management practices while using employee surveys to gather deeper insights.

As future work, a more extensive survey should be conducted, which could give us a more practical overview of RBD in all levels of Cloud System Sustainability, and and by applying the methodology employed in this study, we can define a standardization of sustainable metrics in all the different Risk Levels, including Application, System, System of Systems, and Business Level.

\section*{Acknowledgment}
This work was supported by the CERTAIN project, funded by the
European Union under the Horizon Europe program under Grant Agreement No.\ 101189650.

\bibliographystyle{ieeetr}
\bibliography{bibliography}

\vspace{12pt}

\end{document}